\documentclass[12pt]{article}
\usepackage{amsmath}
\usepackage{graphics}
\usepackage[T2A]{fontenc}%
\usepackage[cp1251]{inputenc}%
\usepackage[english]{babel}%

\begin{document}

\sloppy

\title{Autler-Townes effect in the $D_1$-line hyperfine structure of an alkali atom}

\author{O.S.Mishina, A.S.Sheremet, N.V.Larionov, D.V.Kupriyanov\\%
{\small Polytechnic University 195251, St.-Petersburg, Russia}\\
{\small \rm{E-mail: kupr@dk11578.spb.edu}}}

\maketitle%

\begin{abstract}
We consider the Autler-Townes effect when a strong coupling field is applied in the hyperfine manifold of an alkali atom. Explicit solution is obtained in the case of the D1-line. We show how the hyperfine interaction modifies the dressing effects associated with the strong field as well as the sample susceptibility with respect to the probe mode. Particularly, if the strong field is far detuned from the atomic resonance line the Autler-Townes
structure differs significantly from the prediction of the $\Lambda$-type
approximation. We also find that tuning the strong field in between the upper
state hyperfine components enhances the Autler-Townes effect. The results are
discussed in the context of quantum memory protocols based on the stimulated
Raman process or EIT effect.

 (УДК 535.14, PACS 03.67.Mn, 34.50.Rk, 34.80.Qb, 42.50.Ct)
\end{abstract}

\newpage

\section{Introduction}
A lot of optical phenomena important for the high resolution nonlinear spectroscopy appears as a result of an interaction of several light fields resonant to the coupled transitions in the media \cite{AutlerTownes,LetchChebt}. In the resent years the interest to this problem increased due to the development of the quantum memory systems for light. Quantum state of the field is considered bo be an information carrier \cite{KMP}-\cite{GALS}. Three mechanisms have been studied recently and applied in a practice for the memory: the photon echo effect, the Electromagnetically Induced Transparency (EIT) and the stimulated Raman process. All these three mechanisms are based on an application of the $\Lambda$-type interaction with a strong control field in one arm and a weak probe field in the other arm. The accurate description of this interaction requires a calculation of the interaction line of the atomic system with respect to the probe mode which in modified by the presence of the strong field. This problem was considered in \cite{AutlerTownes} for a two level system dressed by an interaction with the control mode probed by the field on the coupled transition. Authors called it "Stark effect in rapid varying fields". Today this process is known as the Autler-Townes effect. In our work we show the calculation of the Autler-Townes resonance line shape in the hyperfine structure of the $D_1$-line of an alkali atom. It presents an interest for the development of the practical recommendations for the quantum memory system in an optically dense atomic system.

\section{The sample susceptibility}

The level structure of an alkali atom (here ${}^{133}$Cs) is shown in the figure \ref{figone}. Strong monochromatic field with the frequency $\omega$ close to the $D_1$-transition and a weak probe pulse on the coupled transition are presented in the same figure. We assume that only the Zeeman sublevel with the full angular momentum  $F=4$ and its projection $M=4$ is populated to begine with and we call this state $|m\rangle$. The interaction of the control mode in the right circular polarization ($\sigma_+$) with the populated sublevel is excluded completely due to the selection of the $D_1$-line. The strong field interacts with an empty Zeeman sublevel $F=4, M=2$ ($|m'\rangle$ state) in the ground state and with two Zeeman sublevels $F'=3, M'=3$ ($|n\rangle$ state) and $F'=4, M'=3$ ($|n'\rangle$ state) in the excited state.

The presence of the strong mode influences essentially an atomic energy structure. It results in the atomic levels shifts and a distortion of an absorption spectrum in the vicinity of the states $|n\rangle$ and $|n'\rangle$ which is represented in figure \ref{figone} by a dotted line. Also an additional resonance appears in the vicinity of the control field frequency. If the strong mode is tuned on resonance with any of the atomic transitions we observe the splitting of the resonance into two quasi-energy sublevels which are similar in shape. The existence of such a resonance structure can be observed by probing the system "atom + control field" with the field in a left circular polarization ($\sigma_{-}$). This scheme corresponds to the standard method of the Autler-Townes resonance structure observation  \cite{AutlerTownes,LetchChebt} as was mentioned before.

We can separate slowly varying amplitude in a positive frequency component of the probe field
\begin{equation}
{\cal E}^{(+)}_{\mathrm{Left}}(\mathbf{r},t)=%
\epsilon(\mathbf{r}_{\bot},z;t)\mathrm{e}^{-i\bar{\omega}t+i\bar{k}z}%
\label{1}%
\end{equation}
it assumes the propagation of the quasi-monochromatic wave with a carrier frequency $\bar{\omega}$ and a wave number $\bar{k}=\bar{\omega}/c$ in $z$ direction neglecting a diffraction divergence in the transverse plane characterized by the coordinate $\mathbf{r}_{\bot}$. For the Fourier component \begin{equation}
\epsilon(\mathbf{r}_{\bot},z;\Omega)=\int_{-\infty}^{\infty}dt%
\mathrm{e}^{i\Omega t}\epsilon(\mathbf{r}_{\bot},z;t)%
\label{2}
\end{equation}
we can describe the probe field propagation in the sample by the following Maxwell equation
\begin{equation}
\left[-i\frac{\Omega}{c}+\frac{\partial}{\partial z}\right]\!%
\epsilon(\mathbf{r}_{\bot},z;\Omega)%
\!=\!2\pi i\frac{\bar{\omega}}{c}\!%
\tilde{\chi}(\mathbf{r}_{\bot},z;\Omega)\epsilon(\mathbf{r}_{\bot},z;\Omega)%
\label{3}
\end{equation}
Here we introduce the slow varying susceptibility which has the following form in the time domail
\begin{equation}
\tilde{\chi}(\mathbf{r}_{\bot},z;t-t')=\mathrm{e}^{i\bar{\omega}(t-t')}%
\chi(\mathbf{r}_{\bot},z;t-t')%
\label{4}%
\end{equation}
 $\chi(\mathbf{r}_{\bot},z;t-t')$ - it is the standard electrodynamic susceptibility which describes the sample polarization response to the external probe field (\ref{1}). If the sample is probed with the monochromatic field we should consider a spectral dependence of this electrodynamic characteristic of the sample at a frequency $\bar{\omega}$ or a detuning $\bar{\Delta}$: $\chi=\chi(\ldots;\bar{\omega})=\chi(\ldots;\bar{\Delta})$. This value coincide with the value of $\tilde{\chi}(\ldots;\Omega)$ at $\Omega=0$.

The sample susceptibility can be represented as a following expansion
\begin{eqnarray}
\tilde{\chi}(\mathbf{r}_{\bot},z;\Omega)&=&%
-\sum_{n_1=n,n'}\sum_{n_2=n,n'}\frac{1}{\hbar}\left(\mathbf{d e}\right)_{n_1m}^{*}%
\left(\mathbf{d e}\right)_{n_2m}%
\int\frac{d^{3}p}{(2\pi\hbar)^3}n_0(\mathbf{r}_{\bot},z)%
f_0(\mathbf{p})%
\nonumber\\%
&&\times G_{n_1n_2}^{(--)}(\mathbf{p}_{\bot},p_z+\hbar\bar{k};\hbar(\bar{\omega}+\Omega)+\frac{p^2}{2 m})%
\label{5}
\end{eqnarray}
Here we use the matrix elements of an atomic dipole moment operator $\mathbf{d}$ projected to the probe pulse polarization unit vector $\mathbf{e}$; the atom number distribution in space is described by the local concentration $n_0(\mathbf{r}_{\bot},z)$. The distribution of the atoms in momentum space is described by the function $f_0(\mathbf{p})$ which is normalized according to the following condition
\begin{equation}
\int\frac{d^{3}p}{(2\pi\hbar)^3}f_0(\mathbf{p})=1
\label{6}
\end{equation}
The main characteristic which defines the sample susceptibility (\ref{5}) is the retarded Green function of an excited atomic state $G_{n_1n_2}^{(--)}(\mathbf{p},E)$. This functions correspond to all possible indexes combinations of $n_1=n,n'$ and $n_2=n,n'$ and they are presented in the Fourier domain. The energy value $E=\hbar(\bar{\omega}+\Omega)+{p^2}/{2 m_0}$ corresponds to the initial state energy of system "atom + probe field" were the energy of a ground state $|m\rangle$ is taken to be zero. The excited atom momentum is shifted with respect to its momentum in the ground state by the probe mode photon momentum $p_z'=p_z+\hbar\bar{k}$.

We present the Green functions calculated with the use of the diagram technique \cite{MKMP} for a level structure represented in figure \ref{figone}
\begin{eqnarray}
G_{nn}^{(--)}(\mathbf{p},E)&=&\hbar\left\{E-\frac{p^2}{2 m_0}-E_n+i\hbar\frac{\gamma}{2}%
-\frac{|V_{nm'}|^2\left[E-\frac{p^2}{2 m}-E_{n'}+i\hbar\frac{\gamma}{2}\right]}%
{\left[E-E_{n'+}(\mathbf{p},\omega)\right]\left[E-E_{n'-}(\mathbf{p},\omega)\right]}\right\}^{-1}%
\nonumber\\%
G_{n'n}^{(--)}(\mathbf{p},E)&=&\frac{V_{n'm'}V_{nm'}^{*}}%
{\left[E-E_{n'+}(\mathbf{p},\omega)\right]\left[E-E_{n'-}(\mathbf{p},\omega)\right]}%
\times G_{nn}^{(--)}(\mathbf{p},E)%
\nonumber\\%
&&\ldots%
\label{7}%
\end{eqnarray}
The two other functions can be derived by the indexes transposition $n\Leftrightarrow n'$. We introduced the following notations: $V_{nm'}=\left(\mathbf{d
E}\right)_{nm'}$ and $V_{n'm'}=\left(\mathbf{d E}\right)_{n'm'}$ are the matrix elements of the interaction Hamiltonian with respect to the control mode; $\mathbf{E}$ --- complex amplitude of the positive frequency component $\mathbf{E}^{(+)}(\mathbf{r},t)=\mathbf{E}\exp(-i\omega t + i\mathbf{k r})$;
$E_n$, $E_{n'}$ --- unperturbed energies of the atomic excited sublevels; $\gamma$ --- the natural decay rate of an atomic excited state. Quasi-energies $E_{n(n')+}$, $E_{n(n')-}$ from the energy denominators of the equation (\ref{7}) correspond to a "dressing" of the excited hyperfine states independently by the interaction with the control field and the vacuum modes. They are the following
\begin{eqnarray}
E_{n\pm}(\mathbf{p},\omega)&=&E_{m'}+\frac{\mathbf{p}^2}{2m_0}%
+\frac{\hbar}{2}\left[\omega-\frac{\mathbf{k
p}}{m_0}+\omega_{nm'}-i\frac{\gamma}{2}\right]%
\nonumber\\%
&&\pm\left[|V_{nm'}|^2+\frac{\hbar^2}{4}%
\left(\omega_{nm'}-\omega+\frac{\mathbf{k
p}}{m_0}-i\frac{\gamma}{2}\right)^2\right]^{1/2}%
\label{8}%
\end{eqnarray}
and similar for $n'$. To find them we ignore the influence of the hyperfine interaction, and treat each level ($|n\rangle$ or $|n'\rangle$) as if it would be alone and form a pure $\Lambda$-system with the ground state sublevels $|m\rangle$ and $|m'\rangle$. Here $E_{m'}$ and $E_{m}$ are the unperturbed ground state energies which can be taken as a reference point $E_{m'}= E_{m}=0$. The true position of the resonance poles are given by sample susceptibility (\ref{5}) which includes the contributions from all the Green function.

\section{Discussions and results}
\begin{figure}
\center{\includegraphics{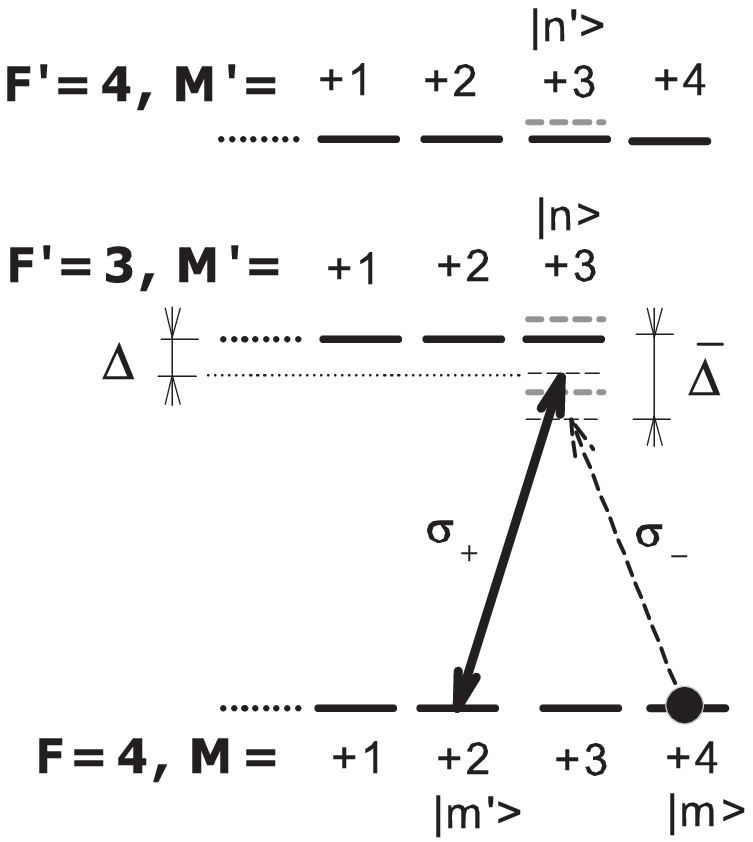}}
\caption{Energy diagram and the excite scheme for the $D_1$-line of ${}^{133}$Cs atom. We consider all the atoms to be on the Zeeman sublevel of the upper hyperfine ground state which is characterized by the maximum projection of the total angular momentum. The system is excited by a strong control field in the right polarization ($\sigma_+$) with the detuning $\Delta$ and it is probed by a weak field in the left polarization ($\sigma_{-}$). By scanning the detuning of the probe field $\bar{\Delta}$ modified resonance structure of an atom "dressed" by the interaction with the control mode is observed which is called the Autler-Townes effect. The position of the quasi-energy states is shown by the gray dotted lines.}
\label{figone}%
\end{figure}

We perform some calculations of the susceptibility $\chi=\chi(\bar{\Delta})=\chi'(\bar{\Delta})+i\chi''(\bar{\Delta})$ for a homogeneous medium. This results are presented in the units of an optical density $n_0({\lambda /{2\pi}})^3$ ($n_0$ - an atomic concentration, $\lambda$ - a carrier wavelength of light) for the different ratio between the control field and a probe field detunings $\Delta$ and $\bar{\Delta}$ (see figure \ref{figone}). We suppose that atoms are placed in a magneto-optic trap at the standard conditions (the temperature is lower than the Doppler limit but higher than the recoil limit). This conditions allow to neglect an effect of an atomic motion. It is important to note that the matrix elements $V_{nm'}$ and $V_{n'm'}$ are no longer independent parameters but strictly related by the rules of the electronic and the nuclear angular momentums summation while considering the hyperfine interaction in the excited atomic state. To describe a coupling with the control field we introduce the Rabi frequency defined by a matrix element which connects the control mode with the low hyperfine sublevel of the excited state $\Omega_c=2|V_{nm'}|/\hbar$. We perform the comparative analysis of our results with the prediction of a $\Lambda$ - scheme model which is neglecting the hyperfine interaction. The $\Lambda$ - scheme assumes  the existence of only one excited sublevel ($|n\rangle=|F'=3,M'=3\rangle$ state) in the $D_1$-line of ${}^{133}$Cs atom.

\begin{figure}
\center{\includegraphics{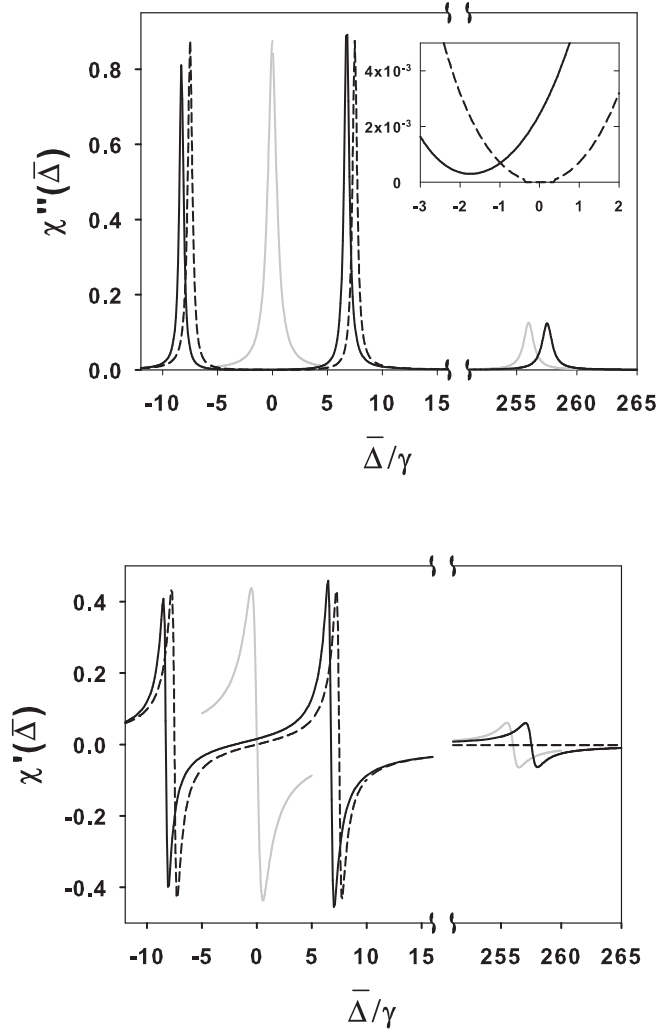}}%
\caption{Absorption (\textit{top}) and a dispersion(\textit{bottom}) components of the sample susceptibility in the case of the resonant excitation $\Delta=0$ and Rabi frequency $\Omega_c=15\gamma$. The solid curve correspond to the presence of the hyperfine interaction and the dotted curve to the $\Lambda$-approach. The gray line correspond to the susceptibility profile in the absence of the control field. In the insert we show that the EIT resonance is shifted to the red wing and the transparency is no longer perfect due to the presence of the second hyperfine sublevel.}
\label{figtwo}%
\end{figure}

\begin{figure}
\center{\includegraphics{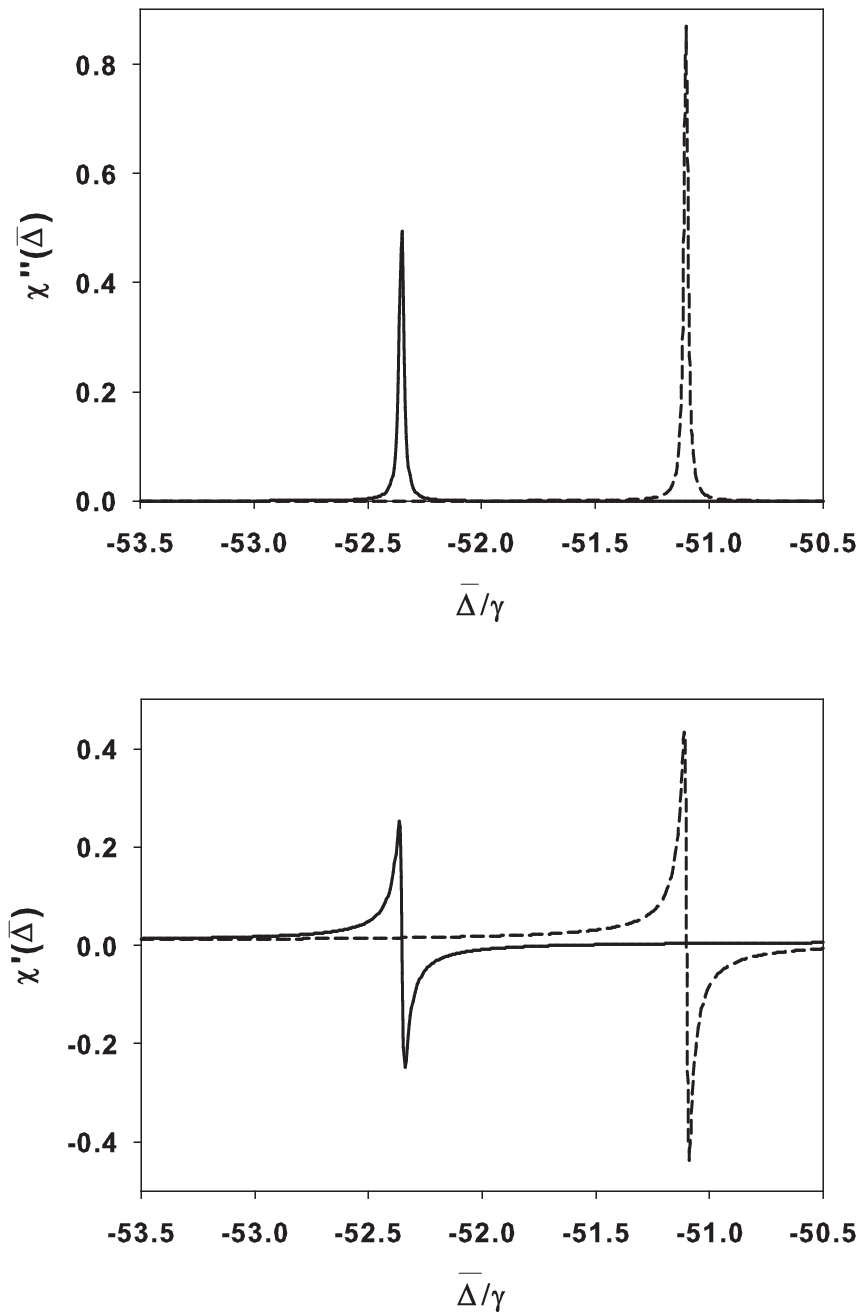}}%
\caption{The same as in figure \ref{figtwo} for the detuning of the control mode $\Delta=-50\gamma$ and Rabi frequency $\Omega_c=15\gamma$. The Autler-Towne's resonance structure is shown in the vicinity of $\bar{\Delta}\sim \Delta$. The structure of the other triplet components is close to the profile of the unperturbed atomic resonance.}
\label{figthree}%
\end{figure}

\begin{figure}
\center{\includegraphics{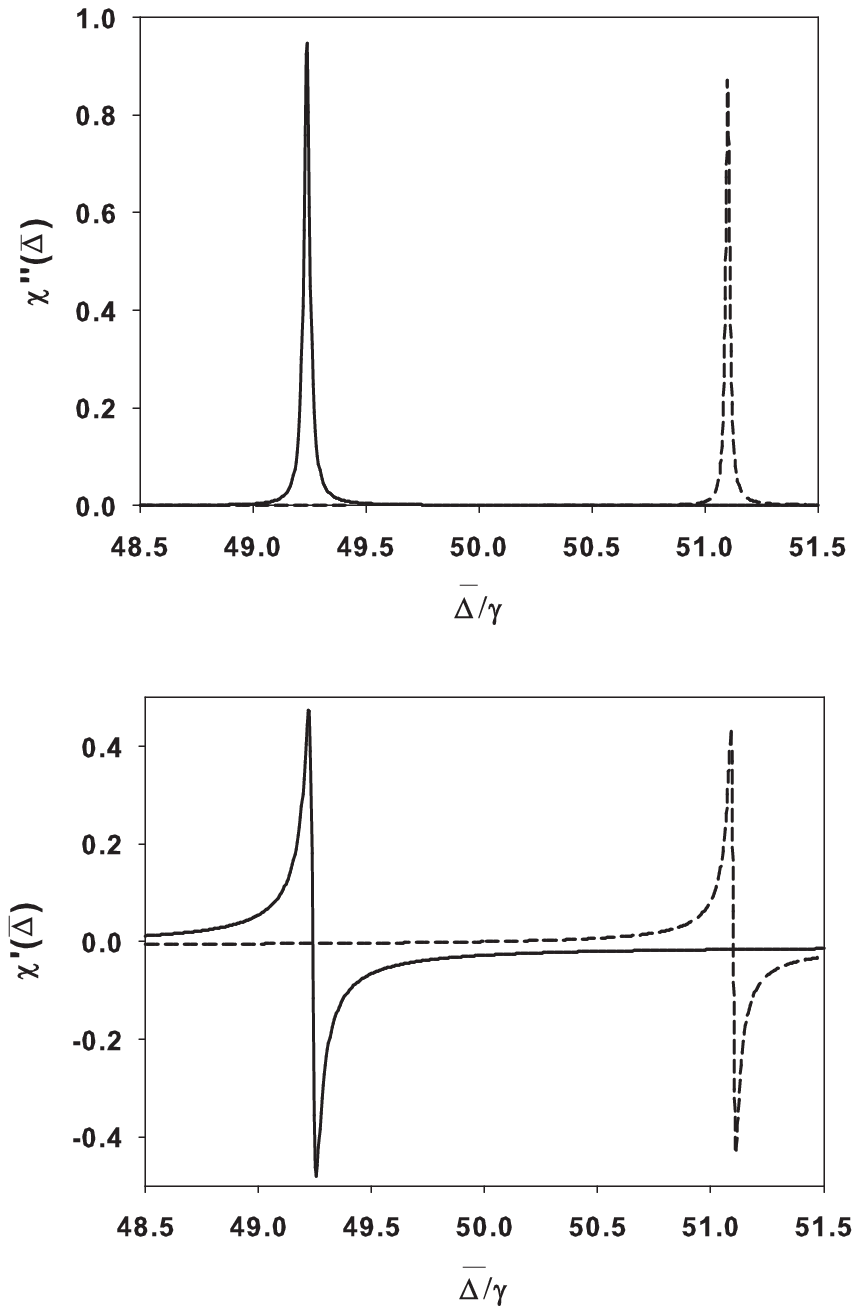}}%
\caption{The same as in figures \ref{figtwo} and \ref{figthree} for the detuning of the control mode $\Delta=+50\gamma$ and Rabi frequency $\Omega_c=15\gamma$. The Autler-Towne's resonance structure is shown in the vicinity $\bar{\Delta}\sim \Delta$. The structure of other triplet components is closed to the profile of the unperturbed atomic resonance.}
\label{figfour}%
\end{figure}

In figures \ref{figtwo}-\ref{figfour} we show an imaginary ("absorption") and a real (dispersion) parts of a sample susceptibility for a similar Rabi frequency $\Omega_c=15\gamma$ and for the different control mode detunings $\Delta$. There is no real absorption in our case but losses in the forward channel due to the incoherent scattering to a solid angel are represented by the imaginary susceptibility component. Figure \ref{figtwo} corresponds to a resonant interaction of the control field with an atomic transition $|m'\rangle\to|n\rangle$. We show all the components of the Autler-Townes triplet in comparison with the three-level model which ignores the state $|n'\rangle$. The difference may look insignificant but in the insert one can see that the hyperfine interaction results to an important qualitative effect. We can see that the ideal conditions for the EIT disappear and the minimum point of the absorption is shifted from the two-photon resonance. The light shift is comparable with natural width of an atomic exited state $\gamma$ and this effect was observed in one of the first slow-light experiment in the optically dense atomic system \cite{LHau}. In the given example the dependence of the $\Lambda$-scheme approach from our model is not so important for the description of the probe pulse delay. Though to use an EIT effect for the quantum memory protocol it is more successful to have  $\Omega_c<\gamma$ \cite{MKMP} when the $\Lambda$-scheme approach also remans.

In figure \ref{figthree} we show the "absorption" and dispersion of the sample susceptibility for the control mode detuning $\Delta=-50\gamma$ as a function of the probe mode detuning in the vicinity of it $\bar{\Delta}\sim -50\gamma$. The result of our calculation using the $\Lambda$-scheme approach with one exited level $|n\rangle$ are shown for the comparison. The essential distinction we observ for the accurate and the approximate calculation is due to the principal role of the hyperfine interaction for the probing of the sample with two orthogonally polarized fields. If the hyperfine interaction dose not take place or it is negligibly small the system would correspond to the atomic transition between the ground state with the anglular momentum $j_0=1/2$ and the excited state with the anglular momentum $j=1/2$ \cite{Happer}. The alignment effect ($\Lambda$ - coupling of Zeeman ground states with $\Delta M=\pm 2$) would not be possible. One can see that the hight of the Autler-Townes resonance in the vicinity of the control mode frequency decreases with the increase of the detuning $\Delta$ to the wing of the $D_1$-line. The calculation for the $\Lambda$ - scheme gives a constant hight of the resonance for the different detunings $\Delta$ while the resonance width is changing.

The situation is changing if the control field frequency is tuned in between the hyperfine components of the excited state as it is shown in a figure \ref{figfour} for $\Delta=+50\gamma$. In this case we can observe an amplification of the effect in comparison with the prediction of $\Lambda$-approach. It is even stronger in the dispersion component of the susceptibility which is more sensitive to the influence of the interference between the hyperfine sublevels, see (\ref{5}) and (\ref{7}). Using such a condition in the optical dense atomic system can lead to a bigger probe pulse delay.This can be used to increase the efficiency of the quantum memory based on a Raman-type storage \cite{KMP,NWRSWWJ,MKMP,GALS}.

\begin{figure}
\center{\includegraphics{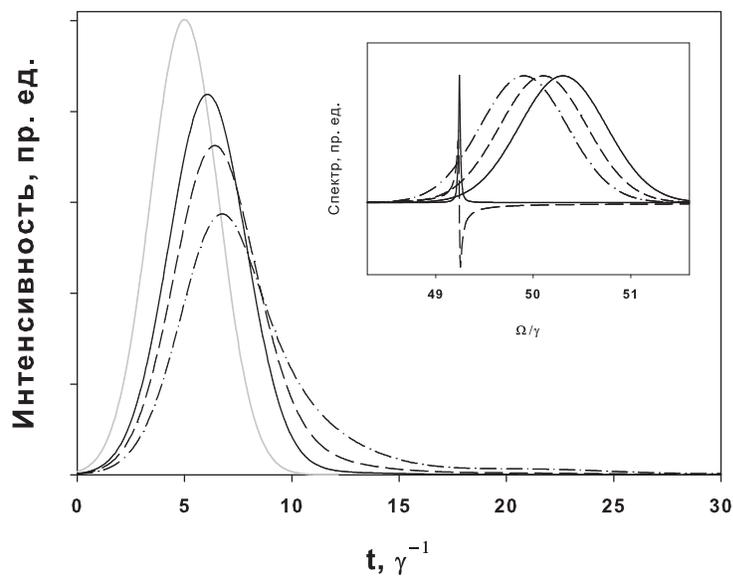}}%
\caption{The dynamics of the probe pulse withe the Gaussian profile passing an opticaly dense atomic sample with the cooperative parameter $n_0({\lambda/{2\pi}})^2L=25$.  This plots correspond to the detuning of the control mode $\Delta=+50\gamma$ and Rabi frequency $\Omega_c=15\gamma$. The input profile is shown by the gray curve and the output pulse profile is changing with respect to the detuning of its carrier frequency. In the insert we show the position of the probe pulse spectrum relatively to the Autler-Townes resonance. When the probe pulse is further detuned from the Autler-Towne's resonance the propagation coefficient is getting close to 90\% and aproximatly 40\% efficiency is achievable for the memory protocol.}
\label{figfive}%
\end{figure}

On the figure \ref{figfive} we demonstrate the delay of a probe pulse which carrier frequency is varied in the vicinity of the Autler-Townes resonance. This resonance corresponds to the control field detuning $\Delta=+50\gamma$ and positioned near the control field frequency. We made our calculation for the homogeneous sample of the length $L$ and the cooperative parameter $n_0({\lambda/{2\pi}})^2L=25$ which is given by the optical depth and is useful to discuss the cooperative effects in one-dimensional systems. It is fully achievable for the experimental conditions in case of an atomic ensemble in a magneto-optical trap with the temperature which is lower than the Doppler limit. One can see that the effect of pulse delay which can be used in order to store the probe light. Storage and retrieval can be done with an efficiency close to 40\% in the system with the transmission coefficient close to 90\%. The retrieval efficiency could be increased by reading with the reversed in space control field which was considered in detail in \cite{GALS}.

\vspace{\baselineskip}%
The materials of this work were reported on the International Conference on Quantum Optics (ICQO'2008) which was organized by the Belarus Science Academy and Lithuania Science Academy. This conference was in Vilnius in September 2008. This work was done with the financial support of INTAS (project №7904) and RFBR-NSF (project №08-02-91355) founds. O.S.M. and A.S.S. thank to the charitable found "Dynasty" for the financial support.

\end{document}